\documentclass[twocolumn,showpacs]{revtex4}
\usepackage{bm}
\usepackage{graphicx}
\usepackage{amsmath}
\usepackage{amssymb}
\usepackage{eufrak}
\usepackage{epsfig}

\newcommand{\nix}[1]{}

\begin{document}

\title{Purcell effect in small metallic cavities}
\author{M.M.~Glazov\footnote{glazov@coherent.ioffe.ru}, E.L.~Ivchenko, A.N.~Poddubny}
\affiliation{A.F.~Ioffe Physico-Technical Institute, Russian
Academy of Sciences, 194021 St.~Petersburg, Russia}
\author{G.~Khitrova}
\affiliation{College of Optical Sciences, The University of Arizona, Tucson, AZ 85721}
\begin{abstract}
We have studied theoretically the Purcell factor which characterizes a change in the emission rate of an electric or magnetic dipole embedded in the center of a spherical cavity. The main attention is paid to the analysis of cavities with radii small  compared to the wavelength. It is shown that the Purcell factor in small metallic cavities varies in a wide range depending on the ratio of the cavity size to the skin depth.
\end{abstract}
  \maketitle
\section{Introduction}
The Purcell effect in the broad sense is defined as a change in the rate of spontaneous emission of a point light source inserted into a resonant cavity. This effect is described by the Purcell factor $f$ defined by the ratio $\tau_{{\rm r},{\rm bulk}}/\tau_{{\rm r},{\rm cav}}$ between lifetimes of spontaneous emission, where $\tau_{{\rm r},{\rm cav}}$ and $\tau_{{\rm r},{\rm bulk}}$ are the lifetimes of the emitting excitation in a system with the cavity and in the infinite homogeneous medium filled with the cavity material. Under the optimal conditions where the resonance frequency of the emitter  $\omega_0$ is tuned to the photon-mode frequency $\omega_c$ and the emitter is placed in the field antinode the Purcell factor is given by~\cite{Purcell}
\begin{equation} \label{Pfactor}
f = \frac{3Q \lambda^3}{4 \pi^2 V}\:.
\end{equation}
Here $V$ is the volume of the resonator, $Q$ is the quality factor, $\lambda \equiv \lambda_0$ is the light wavelength in vacuum if the cavity is empty and $\lambda = \lambda_0/n$ if the cavity is filled by a substance with the refractive index $n$. A brief communication~\cite{Purcell} on spontaneous emission of the oscillating (nuclear) magnetic dipole contains three estimations of the factor $f$. For a resonant metal cavity with the linear dimension $a$ and the metal skin depth $\delta$, the factor $f$ is 
\begin{equation} \label{purcell1}
f \sim \frac{\lambda^3}{a^2 \delta}\:.
\end{equation}
For nonresonant systems, $\lambda \ll a$, Ref.~[\onlinecite{Purcell}] presents the estimation formulae
\begin{equation} \label{purcell23}
f \sim \frac{\lambda^3}{a^3}\hspace{5 mm} \mbox{and} \hspace{5 mm} f \sim \frac{\lambda^3}{a \delta^2}\:,
\end{equation}
the latter as applied to the case $a < \delta$. In contrast to the well-known and widely cited Eq.~(\ref{Pfactor}), see e.g. the review \cite{oraevskii}, the above two estimations, as far as we know, are not referred to in the literature. In the present work we have calculated the Purcell factor $f$ for the emitter placed into a nonresonant spherical metal cavity of radius $a \ll \lambda$ and obtained the following results for the emitting magnetic dipole 
\begin{equation}
\label{thinthick:magnetic2}
f = \left\{ \begin{array}{c}  9\delta/(2k_1^3 a^4)\:, \hspace{4 mm} \mbox{if} \hspace{6 mm} a \gg \delta\:,\\
 2/(\delta^2 k_1^3 a) \:, \hspace{6 mm} \mbox{if} \hspace{6 mm} a \ll \delta\:, \end{array} \right. 
\end{equation}
and the emitting electric dipole 
\begin{equation}
\label{thinthick}
 f= \frac98\left\{ \begin{array}{c}  \delta/(k_1 a^2)\:, \hspace{6 mm} \mbox{if} \hspace{5 mm} a \gg \delta\:,\\
 \delta^2/(k_1 a^3) \:, \hspace{5 mm} \mbox{if} \hspace{5 mm} a \ll \delta\:, \end{array} \right. 
\end{equation}
where $k_1 = 2 \pi/\lambda$. Comparison of Eqs.~(\ref{purcell23}) and Eqs.~(\ref{thinthick:magnetic2}) shows agreement with \cite{Purcell} for the magnetic-dipole emission in the particular case $a \ll \delta$; on the other hand, the estimation $f\sim (\lambda/a)^3$ corresponds (as shown below) to the maximum possible value of Purcell factor for an ultrasmall cavity with a magnetic dipole at $\delta\sim a$. 

For obtaining formulae (\ref{thinthick:magnetic2}) and (\ref{thinthick}) we will derive a general expression for the factor $f$ for arbitrary values of $\lambda, a$ and dielectric permittivities $\varepsilon_1, \varepsilon_2$ of the materials inside and outside the sphere of radius $a$, respectively. Then we will consider various particular cases, including those where Eqs.~(\ref{thinthick:magnetic2}), (\ref{thinthick}) are valid. In what follows we assume the  permittivity $\varepsilon_1$ to be real whereas no restrictions are imposed on the real and imaginary parts of $\varepsilon_2$ (except for the natural condition ${\rm Im}\, \varepsilon_2 \geq 0$). 
\section{Emission of an electric dipole}
For the sake of definiteness we consider a spherical semiconductor quantum dot placed in the center of a spherical cavity.  The notations $\varepsilon_1$ and $\varepsilon_2$ are used for the  permittivities of the cavity material and the surrounding medium, respectively. The difference between $\varepsilon_1$ and the background dielectric constant of the quantum dot is neglected. The quantum dot radius $r_{\rm QD}$ is set to be small compared with the cavity radius $a$. The electric field of the light wave emitted by the quantum dot satisfies the wave equation
\[
{\rm rot} ~ {\rm rot}~ {\bm E}({\bm r}) - k_0^2 \varepsilon({\bm r}) {\bm E}({\bm r}) = 4 \pi k_0^2 {\bm P}_{\rm exc}({\bm r})\:,
\]
or, equivalently,
\begin{multline} \label{MaxwE}
\Delta {\bm E}({\bm r}) + k_0^2 \varepsilon({\bm r}) {\bm E}({\bm r}) =
\\
 - 4 \pi k_0^2 [{\bm P}_{\rm exc}({\bm r}) + 
k_1^{-2} {\rm grad}~{\rm div}~ {\bm P}_{\rm exc}({\bm r})]\:.
\end{multline}
Here \[
\varepsilon({\bm r}) = \left\{ \begin{array}{c} \varepsilon_1 \hspace{4 mm} \mbox{for} \hspace{2 mm}
r<a \:, \\  \varepsilon_2 \hspace{4 mm} \mbox{for} \hspace{2 mm} r > a\:, \end{array} \right.
\]$
k_0 = \omega/c$, $\omega$ is the (complex) eigenfrequency of the exciton excited in the quantum dot; it is determined from 
the closed algebraic equation derived with allowance for the exciton-photon coupling (see below).
Here $\bm P_{\rm exc}({\bf r})$ is the contribution of the selected exciton resonance to the dielectric polarization given by~\cite{Ivch_1992,Ivchenkobook2}
\begin{equation}
\label{polarization}
4 \pi \bm P_{\rm exc}({\bf r})=\frac{\pi \varepsilon_1 \omega_{LT} a_B^3}{\omega_0 - \omega - {\rm i} \Gamma} \Phi({\bm r})\int d^3 r'\Phi({\bm r}')\bm E({\bf r}')\:,
\end{equation}
$\omega_0$ is the bare exciton resonance frequency (not renormalized by the exciton-photon coupling), $\Gamma$ is the nonradiative exciton damping rate (in the following neglected, for simplicity), $a_B$ and $\omega_{LT}$ are the Bohr radius and longitudinal-transverse splitting of the exciton in the bulk semiconductor, $\Phi({\bm r})$ is the envelope of the exciton wave function at coinciding electron and hole coordinates, it is assumed to be isotropic: $\Phi({\bm r}) \equiv \Phi(r)$.  A schematic illustration of the system under consideration is presented in the inset in Fig.~1.

Let us decompose the complex eigenfrequency into real and imaginary parts, namely, $\omega = \tilde{\omega}_0 - {\rm i} \Gamma_0$, where $\tilde{\omega}_0 - \omega_0$ is the renormalization of the resonance frequency due to the electron-photon coupling, and the exciton radiative lifetime $\tau_r$ is related to the radiative damping rate by $\tau_r^{-1} = 2 \Gamma_0$. In order to find $\tilde{\omega}_0$ and $\Gamma_0$ we apply the tensor Green function satisfying the wave equation
\cite{Ivch_1992,Kalit,ajiki}
\begin{equation} \label{Greenfunction}
[\Delta + k_0^2 \varepsilon(\bm r) ]~G_{\alpha \beta}({\bm r}, {\bf r}') = - \left(\delta_{\alpha \beta} + \frac{1}{k_1^2} \frac{\partial^2}{\partial r_{\alpha} \partial r_{\beta}} \right) \delta(\bm r  - {\bf r}')\:.
\end{equation}
Taking into consideration the spherical symmetry of the cavity with the quantum dot placed in the center we obtain
\begin{multline} \label{omega0gamma0}
\tilde{\omega}_0 - \omega_0 - {\rm i} \Gamma_0 = \\
- \pi k_1^2 a_B^3 \omega_{LT} \iint d^3r d^3r' \Phi(r)\Phi(r')~G_{xx}(\bm r, \bm r') \:, 
\end{multline}
where $k_1 = \sqrt{\varepsilon_1} k_0$. Equation (\ref{omega0gamma0}) is valid for the weak exciton-photon coupling in which case the argument $\omega$ of the Green function can be replaced by the bare exciton frequency $\omega_0$.  In the strong exciton-photon coupling regime, one has to make allowance for the explicit dependence of the Green function on the frequency  $\omega$ and solve an algebraic equation for the eigenfrequencies of the zero-dimensional exciton polaritons \cite{Kalit}.

The present work is aimed at the calculation of the exciton radiative lifetime in the regime of weak coupling of the exciton with the electromagnetic radiation. Therefore, we retain in the left- and right-hand sides of Eq.~(\ref{omega0gamma0}) only imaginary parts and present the general equation for the radiative damping rate
\begin{multline} \label{gamma0}
\Gamma_0(\varepsilon_1, \varepsilon_2, a) = 
\\
\pi k_1^2 a_B^3 \omega_{LT} \iint d^3r d^3r' \Phi(r)\Phi(r')~{\rm Im}\{G_{xx}(\bm r, \bm r')\}\:. 
\end{multline}
We introduced the variables $\varepsilon_1, \varepsilon_2$ and $a$ in the notation of the exciton damping rate, just as a reminder. 

By using an explicit expression for the Green function \cite{Ivchenkobook2}, we can transfer Eq.~(\ref{gamma0}) to
\begin{equation} \label{gamma0gamma0}
\Gamma_0(\varepsilon_1, \varepsilon_2, a) = f \Gamma_0(\varepsilon_1)\:, 
\end{equation}
where $\Gamma_0(\varepsilon_1)$ is the exciton radiative damping rate in the homogeneous medium with dielectric constant $\varepsilon_1$, see \cite{Ivch_1992}:
\begin{eqnarray} \label{gamma0hom}
\Gamma_0(\varepsilon_1) &=& \frac{1}{6} k_1^2 a_B^3 \omega_{LT} \iint d^3r d^3r' \Phi(r)\Phi(r')\frac{\sin{k_1|{\bm r} - {\bm r}'|}}{|{\bm r} - {\bm r}'|} \nonumber
\\ &=& \frac{1}{6}k_1^3 a_B^3\omega_{\rm LT}\left[\int d^3r \frac{\sin{k_1 r}}{k_1 r} \Phi(r) \right]^2\:, 
\end{eqnarray}
and the Purcell factor is related by 
\begin{equation} \label{purcell_factor}
f = 1 + {\rm Re}~R^{\rm TM}_{12,1}
\end{equation}
with the ``reflection'' coefficient $R^{\rm TM}_{12,l}(\omega)$ of the electric-dipole (TM) light wave with the total angular momentum $l=1$ at the frequency $\omega=\omega_0$. According to \cite{ajiki} the explicit form of $R^{\rm TM}_{12,1}(\omega)$ reads
\begin{equation} \label{rtm121}
R^{\rm TM}_{12, 1} = \frac{\sqrt{\varepsilon_2} \xi_1(k_2 a) \xi^{\prime}_1(k_1a) -
\sqrt{\varepsilon_1} \xi^{\prime}_1(k_2a)\xi_1(k_1a) }{\sqrt{\varepsilon_1} \psi_1(k_1a) \xi^{\prime}_1(k_2a) -
\sqrt{\varepsilon_2} \psi^{\prime}_1(k_1a)\xi_1(k_2a)}\:.
\end{equation}
Here we use the notations $\psi_1(x) = x j_1(x)$, $\xi_1(x) = x h^{(1)}_1(x)$, where $j_1(x)$ is the spherical Bessel function, $h^{(1)}_1(x) = j_1(x) + {\rm i} y_1(x)$ is the spherical Hankel function, the prime indicates differentiation over its variable $x$. These functions satisfy the identities
\begin{eqnarray}\label{bessel}
&&[x h^{(1)}_1(x)]^{\prime} = x h^{(1)}_0(x) - h^{(1)}_1(x)\:,\:
[x j_1(x)]^{\prime} = x j_0(x) - j_1(x)\:, \nonumber\\
&&  \hspace{ 15 mm} j_0(x) = \frac{\sin{x}}{x}\:,\:j_1(x) = \frac{\sin{x}}{x^2} - \frac{\cos{x}}{x}\:,  \nonumber\\
&&  \hspace{ 15 mm} y_0(x) =  - \frac{ \cos{x} }{x}\:,\:
y_1(x) = - \frac{\cos{x}}{x^2} - \frac{\sin{x}}{x} \:. \nonumber
\end{eqnarray}
By virtue of these identities Eq.~(\ref{purcell_factor}) yields the expression 
\begin{equation} \label{fRTM}
f =   {\rm Im} \left[ \frac{( 1 + S_d) y_1(k_1a) -  y_0(k_1a) k_1 a }{(1 + S_d) j_1(k_1a) -  j_0(k_1 a) k_1a} \right]\:
\end{equation}
with
\begin{equation} \label{Sfactor}
S_d = \frac{k_1^2 a \xi^{\prime}_1(k_2a)}{k_2\xi_1(k_2a)} = 
\frac{k_1^2}{k_2^2} \left( - 1 + \frac{(k_2a)^2}{ 1 - {\rm i} k_2a} \right),
\end{equation}
for the Purcell factor convenient for the further analysis.
Multiplying the numerator and denominator in (\ref{fRTM}) by the complex conjugate denominator, taking into account that the wave number $k_1$ is real and $j_1(x)y_0(x)-y_1(x)j_0(x)=1/x^2$ we obtain
\begin{equation} \label{fRTM2}
f =\frac{{\rm Im}\{
S_d\}}{k_1a|(1 + S_d) j_1(k_1a) -  j_0(k_1 a) k_1a|^2}\ . 
\end{equation}
In the following subsections we apply the general equations (\ref{fRTM}) and (\ref{fRTM2}) for finding the Purcell factor in several particular cases.
\subsection{Spherical resonant microcavity}
Strictly speaking, the cavity volume $V = (4 \pi/3)a^3$ in Eq.~(\ref{Pfactor}) should be replaced by the effective volume $\tilde{V}$ different from $V$ because the electric (magnetic) field inside the cavity is inhomogeneous and the Purcell factor is determined by the field enhancement in the point of the emitter location \cite{coccioli,andreani,robinson,koenderink}. 

Let us show that the Purcell equation (\ref{Pfactor}) follows from Eq.~(\ref{fRTM}) and determine the ratio $V/\tilde{V}$
for the main TM mode (for the TE mode this ratio is presented in the next section). The photon modes in a spherical cavity are found from the minimum condition for the modulus of the denominator in the right-hand side of Eq.~(\ref{fRTM}). Let us introduce the function 
\[
F(x) = \frac{xj_0(x)}{j_1(x)} = \frac{x^2}{1 - x \cot{x}}\:.
\]
In the structure satisfying the condition $|S_d| \ll 1$, the size-confined TM modes in the zeroth order in $S_d$ are found  from the equation $F(k_1 a) = 1$. The first root of this equation equals 2.7437, see \cite{IEEE,ll8}. We introduce the notations $\omega^*$ and $k_1^* = \sqrt{\varepsilon_1} \omega^*/c$ for the frequency and wave number of this mode. The exact value of the complex eigenfrequency $\omega = \tilde{\omega}^* - {\rm i} \gamma^*$ is extracted from the equation
\[
F\left( \frac{\sqrt{\varepsilon_1}\omega a}{c} \right) - 1 = S_d(\omega)\:.
\]
A correction of the first order in $S_d$ can be found by retaining linear terms in the expansion of $F$ in powers of $\omega - \omega^*$ and replacing $S_d(\omega)$ by $S_d(\omega^*)$. The result yields
\[
\tilde{\omega}^* = \omega^* - \frac{c\; {\rm Re}[ S_d(\omega^*) ]}{\sqrt{\varepsilon_1}a \left\vert F'(k_1^*a) \right\vert} \:,
\:\gamma^* = \frac{c\; {\rm Im}[ S_d(\omega^*) ]}{\sqrt{\varepsilon_1}a \left\vert F'(k_1^*a) \right\vert}  \:.
\]
The quality factor $Q$ (which is the standard parameter of a resonator) is related with the photon mode damping rate $\gamma^*$ by
$\gamma^* = \tilde{\omega}^*/(2Q)$. Since, for any root $x^*$ of equation $F(x)=1$, one has $F'(x^*) = (2 - x^{* 2})/x^*$, the exciton damping rate  is given by
\[
\Gamma_0(\varepsilon_1, \varepsilon_2, a) = \left( \frac{\Delta}{2} \right)^2 
\frac{\gamma^*}{(\omega_0 - \tilde{\omega}^*)^2 + \gamma^{* 2}}\:,
\]
where we introduced the Rabi splitting
\[
\Delta = 2 \sqrt{ \frac{c \Gamma_0(\varepsilon_1)}{\sqrt{\varepsilon_1}a}
\frac{1 - x^{*2} + x^{*4}}{x^{*2} |2 - x^{*2}|}}\:.
\]
In the weak-coupling regime the inequality $\Delta \ll 2 \gamma^*$ is satisfied (we remind that we consider the case of negligibly small nonradiative damping rate,  $\Gamma \ll \gamma^*$). Under exact resonance condition $\omega_0 =\tilde{\omega}^*$ we obtain for the Purcell factor
\begin{equation} \label{fres}
f = \frac{3Q \lambda_1^3}{4 \pi^2 \tilde{V}}\:,\quad \frac{V}{\tilde{V}} = \frac49~ \frac{1 - x^{*2} + x^{*4}}{|2 - x^{*2}|}\:.
\end{equation}
For $x^* = 2.7437$, the ratio $V/\tilde{V} \approx 4$.  Note that, in the resonant cavities, the three quantities $V$, $\tilde V$ and $\lambda_1^3$ differ only by numerical factors. Therefore, the Purcell factor coincides by the order of magnitude with the quality factor. Indeed, the electric-field amplitude is enhanced in the resonator to the extent of high quality factor $Q$.

Now we turn to the nonresonant cavities of ultrasmall size satisfying the condition $k_1 a~\ll~1$.
\subsection{Small nonresonant cavity in nonabsorbing media}
Here we assume that the dielectric constants $\varepsilon_1$, $\varepsilon_2$ are real and positive. Taking into account the following expansions for small values of the variable
\begin{equation} \label{mathem}
x j_0(x) \approx x\:,\: j_1(x) \approx \frac{x}{3} 
\end{equation}
and substituting them into Eq.~(\ref{fRTM2}) we obtain
\begin{equation} \label{smallcavity}
f = \frac{9\; {\rm Im} [ S_d(\omega_0)]}{(k_1a)^3 |2-S_d|^2} \:.
\end{equation}

In the particular case $k_2a \ll 1$ one has
\[
{\rm Im}~S_d \approx k_1^2k_2 a^3 \ll 1\:,\: {\rm Re}\;S_d \approx - \frac{\varepsilon_1}{\varepsilon_2}
\]
and the Purcell factor is given by the well-known expression \cite{ohtaka,glauber,Lavallard,Yablon,khitrova1,GoupPRB}
\begin{equation} \label{diel1}
f = \sqrt{ \frac{\varepsilon_2}{\varepsilon_1} } \left( \frac{3 \varepsilon_2}{2 \varepsilon_2 + \varepsilon_1} \right)^2\:.
\end{equation}
Qualitatively, Eq.~\eqref{diel1} can be interpreted as a classical enhancement of the stationary electric field in the dielectric cavity. 

In the opposite limiting case $k_2a \gg 1$ the Purcell factor equals to
\begin{equation} \label{diel2}
f = \frac{9}{4k_1 k_2 a^2}
\end{equation}
and can be both larger and smaller than unity.
\subsection{Nonresonant cavity made of metal}
Next we consider a small metallic cavity with $k_1a \ll 1$ and the  permittivity
\begin{equation} \label{sigma}
\varepsilon_2(\omega) = 1+\frac{4 \pi {\rm i} \sigma}{\omega} = 1+\frac{2 {\rm i}}{\delta^2} \left( \frac{c}{\omega}\right)^2\:,
\end{equation}
where $\sigma$ and $\delta$ are the metal static conductivity and skin depth, respectively. In the quasi-stationary approximation, i.e., for $\delta \ll c/\omega$, one can neglect unity in Eq.~\eqref{sigma} reducing this equation to
\begin{equation} \label{sigma2}
\varepsilon_2(\omega) = \frac{2 {\rm i}}{\delta^2} \left( \frac{c}{\omega}\right)^2\:,\:
  k_2 = \frac{1 + {\rm i}}{\delta}\:.
\end{equation}
It follows from Eqs.~(\ref{Sfactor}), (\ref{smallcavity}) that for a thin skin depth, i.e., for $\delta \ll a$ so that $|k_2a| \gg 1$, one has 
\[
S_d \approx {\rm i} \frac{k_1^2a}{k_2}\:,\: {\rm Im}\{ S_d \} = \frac12 k_1^2 a \delta \hspace{3 mm} \mbox{and}
\hspace{3 mm} |S_d| \ll 1\:,
\]
which leads to the first Eq.~(\ref{thinthick}). If the skin depth exceeds the linear dimension of the cavity but is small as compared with the wavelength so that $\delta \gg a$ and $k_1 \delta \ll 1$, one has $|S_d| \ll 1$ and ${\rm Im}\{ S_d \} = - {\rm Im}\{ k_1^2/k_2^2 \}$ = $k_1^2 \delta^2/2$ which leads to the second equation (\ref{thinthick}). 
In the case where the skin depth is the longest among $\delta, a$ and $1/k_1$ the quasi-stationary approximation is invalid and the wave vector $k_2$ must be determined from the equation~(\ref{sigma}) for $\varepsilon_2$. In this regime the following asymptotics for the Purcell factor holds
\begin{equation}\label{very_large_delta}
 f=\frac{18\varepsilon_1^2}{\delta^2k_1^5a^3(\varepsilon_1+2)^2}+
\frac{9}{\sqrt{\varepsilon_1}(2 + \varepsilon_1)^2}\:.
\end{equation}
In the limit $\delta\to\infty$ the Purcell factor is given by the second term in Eq.~(\ref{very_large_delta}).
In this case the metal is transparent,  $\varepsilon_2\equiv 1$, and the second term in Eq.~(\ref{very_large_delta}) is equivalent to the expression (\ref{diel1}) from the previous section.

\begin{figure}[t]
\includegraphics[width=\linewidth]{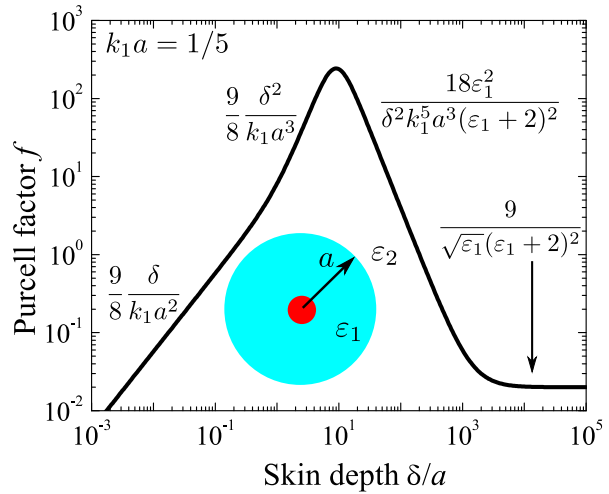}
\caption{Dependence of the Purcell factor $f$ on the skin depth~$\delta$ in the metallic cavity. The calculation is performed by the general equation~\eqref{fRTM2} for $k_1a =1/5$ and $\varepsilon_1=10$. Approximate analytical dependences $f(\delta)$ are also indicated. Inset shows schematics of the structure under consideration: quantum dot (inner sphere) embedded in the center of the spherical medium with  permittivity $\varepsilon_1$ (outer sphere) which, in its turn, is surrounded by the medium with the permittivity $\varepsilon_2$.}\label{fig:stokes}
\end{figure}

Figure 1 presents the calculated dependence of the Purcell factor $f$ on the skin depth $\delta$ expressed in units of the cavity radius $a$. The curve depicted in the double logarithmic scale demonstrates three regions in the dependence $f(\delta)$ where it is described by power-law dependence indicated in the figure. The first two regions of the curve corresponding to small values of $\delta \ll 1/k_1$ demonstrate linear and quadratic dependences of the Purcell factor on $\delta$, in agreement with Eq.~\eqref{thinthick}. At $\delta\sim 1/ k_1$ the factor reaches a maximum value of $f\sim 1/(k_1 a)^3$ (in this estimation the numerical factor dependent on $\varepsilon_1$ is omitted) and then, at $\delta \gg 1/k_1$, decreases as $1/\delta^2$. Finally, in region of $\delta$ where the criterion for quasi-stationary approximation is violated and $\delta\gg  a(k_1a)^{-5/2}$, the Purcell factor saturates to a constant value described by the second term in Eq.~(\ref{very_large_delta}). It should be added, however, that at high frequencies the approximate equation (\ref{sigma}) for the permittivity of metal must be replaced by
\[
\varepsilon_2(\omega) = 1 - \frac{\omega^2_{\rm pl}}{\omega(\omega + {\rm i} \gamma_m)} \:,
\]
where $\omega_{\rm pl}$ is the plasma frequency and $\gamma_m$ is the 
damping rate describing the relaxation of electrons in the metal. If that is the case, the above expression for $\varepsilon_2(\omega)$ can be substituted into Eq.~(\ref{fRTM2}) to calculate the Purcell factor more precisely.

Qualitatively the obtained results can be interpreted as follows. For small values of skin depth the penetration length of the electric field from the cavity into the metal is small. Therefore, the spontaneous emission of the quantum dot is suppressed in the limit $\delta \to 0$ and intensifies with increasing $\delta$. In the opposite limit $\delta\to \infty$ where $\varepsilon_2 \to 1$, the electric field inside the cavity with $\varepsilon_1>1$ is smaller than outside which leads to weakening of the spontaneous emission. At the intermediate values of $\delta$ a remarkable penetration of the field into the metal results in an increase of the Purcell effect. This enhancement of the spontaneous decay is ensured by the efficient absorption (in the metal) of the wave radiated by the quantum dot. 

\section{Emission of a magnetic dipole}
Let the system with a spherical cavity be as before characterized by the dielectric permittivities $\varepsilon_1$ and $\varepsilon_2$ but, instead of the electric dipole, a magnetic dipole be placed in the cavity center. Again, we assume that the magnetic dipole is distributed with the density ${\bm M}_{\rm ext}({\bm r})$ in the small region of radius $r_{\rm QD} \ll a$ and oscillates at the resonant frequency $\omega_0$. Then, instead of Eq.~(\ref{MaxwE}) for the electric field ${\bm E}$, it is more convenient to solve the wave equation for the magnetic field 
\begin{multline} \label{MaxwH}
\Delta {\bm H}({\bm r}) + k_0^2 \varepsilon({\bm r}) {\bm H}({\bm r}) =\\
 - 4 \pi k_1^2 [{\bm M}_{\rm ext}({\bm r}) + 
k_1^{-2} {\rm grad}~{\rm div}~ {\bm M}_{\rm ext}({\bm r})]\:.
\end{multline}
The solution for ${\bm H}({\bm r})$ can be written by using the same Green function (\ref{Greenfunction}). 
Note that, as compared with Eq.~(\ref{MaxwE}), here the right-hand side contains the factor $k_1^2 = k_0^2 \varepsilon_1$, rather than $k_0^2$. Moreover, in the magnetic-dipole case, in the expansion of the Green function over spherical harmonics \cite{ajiki}, one should take into account the TE-wave with the total angular momentum $l=1$ which has an antinode of the magnetic field in the cavity center. As a result, the Purcell factor equals to $1 + {\rm Re}~R^{\rm TE}_{12,1}$, where \cite{ajiki} 
\[
R^{\rm TE}_{12, 1} = \frac{\sqrt{\varepsilon_1} \xi_1(\kappa_2) \xi^{\prime}_1(\kappa_1) -
\sqrt{\varepsilon_2} \xi^{\prime}_1(\kappa_2)\xi_1(\kappa_1) }{\sqrt{\varepsilon_2} \psi_1(\kappa_1) \xi^{\prime}_1(\kappa_2) -
\sqrt{\varepsilon_1} \psi^{\prime}_1(\kappa_1)\xi_1(\kappa_2)}\:.
\]
By using the properties of spherical Bessel functions we can reduce the expression for $f$ to a form similar to Eq.~(\ref{fRTM}), namely, 
\begin{equation} \label{fmTE}
f =  1 + {\rm Re}~R^{\rm TE}_{12,1} = {\rm Im} \left[ \frac{( 1 + S_m) y_1(k_1a) - k_1a  y_0(k_1a)}{(1 + S_m) j_1(k_1a) - k_1a j_0(k_1a)} \right] \:,
\end{equation}
where
\[
S_m =  \frac{k_2^2}{k_1^2}~ S_d =  - 1 + \frac{(k_2a)^2}{ 1 - {\rm i} k_2a} \:.
\]
Instead of Eq.~\ref{fRTM2}), we obtain
\begin{equation} \label{fRTE2}
f = \frac{~ {\rm Im}\{
S_m\}}{k_1a|(1 + S_m) j_1(k_1a) -  j_0(k_1 a) k_1a|^2}\:.
\end{equation}
In the nonresonant metal cavity satisfying the condition $k_1 a \ll 1$, one can apply Eq.~(\ref{smallcavity}) where $S_d$ is replaced by $S_m$. For a small dielectric cavity with $k_1 a \ll 1$ the following asymptotics are valid : $S_m \approx {\rm i} k_2 a$ for $|k_2|a \gg 1$ and $S_m \approx -1 + (k_2 a)^2$ for $|k_2|a \ll 1$. As a result, we obtain instead of Eqs.~(\ref{diel1}) and (\ref{diel2}): $f = (k_2/k_1)^3 = (\varepsilon_2/\varepsilon_1)^{3/2}$ for $|k_2|a \ll 1$ and $f = 9/(k_1^3 k_2 a^4)$ for $|k_2|a \gg 1$.

In metallic cavities with the response described by Eq.~\eqref{sigma} and for $|k_2|a \ll 1$, we come to Eq.~(\ref{thinthick:magnetic2}). For arbitrary relation between $\delta$ and $a$ but still for $k_1 a \ll 1$, the Purcell factor is described by
\begin{equation}
 \label{purcell:magn:gen}
f=\frac{18\delta(\delta+a)}{k_1^3a(4a^4+12a^3\delta+18a^2\delta^2+18a\delta^3+9\delta^4)}\:.
\end{equation}
As well as for an electric dipole, the dependence of the Purcell factor on the skin depth (for a fixed geometry of the cavity)
is nonmonotonous; a distinctive feature is that the Purcell factor reaches a maximum at $\delta \sim a \ll 1/k_1$. This can be readily checked from the general equation \eqref{purcell:magn:gen} or from the estimations (\ref{thinthick:magnetic2}). Hence by the order of magnitude the maximum enhancement of magnetic-dipole emission amounts to $f \sim 1/(k_1 a)^3$. 

Now we will briefly analyze the resonant systems containing magnetic dipoles. A high-quality cavity satisfies the condition
$|k_2| a \gg 1$. In this case $S_m \approx {\rm i} k_2 a$ and, in the zeroth order in the small parameter $|S_m|^{-1} \ll 1$,  TE eigenmodes are found from the equation $j_1(k_1a) = 0$, or $\tan{k_1 a} = k_1a$. The lowest root of the equation $j_1(y) = 0$ is $y^* = 4.4934$, in agreement with \cite{IEEE,ll8}, and the lowest frequency of the TM mode equals $\omega_m^* \equiv y^* c /( \sqrt{\varepsilon_1} a)$. 

For close resonance frequencies of the emitter and TE mode, i.e., at $\omega_0 \approx \omega_m^*$, the inequality $|S_m| \gg 1$ allows one to reduce the expression (\ref{fmTE}) to 
\[
f = {\rm Im} \left[ \frac{y_1(y^*)}{ j'_1(y^*) (k_1a - y^*) - y^* j_0(y^*) S_m^{-1}} \right]
\]
\[
= \frac{1 + y^{*2}}{y^{*2}} \frac{c}{a \sqrt{\varepsilon_1}}  \frac{\gamma_m^*}{ ( \omega_0 - \tilde{\omega}_m^*)^2 
+ \gamma_m^{*2} }\:,
\]
where
\[
\tilde{\omega}^* - \omega^* = - {\rm Im}\, \frac{\omega_m^*}{k_2a}\:,\: \gamma_m^* = {\rm Re}\, \frac{\omega_m^*}{k_2a}\:.
\]
The quality factor is given by $Q_m = [{\rm Re}(2/k_2a)]^{-1} = \delta/a$, in agreement with Eqs.~(\ref{Pfactor}) and (\ref{purcell1}). In the system tuned to the exact resonance, $\omega_0 = \tilde{\omega}_m^*$, the Purcell factor is determined by the formula (\ref{fres}) with the ratio of volumes
\[
\frac{V}{\tilde{V}} = \frac49 (y^{*2} + 1)\:.
\]

\section{Analysis of the field structure and discussion}
According to Eq.~(\ref{purcell_factor}) the Purcell factor is determined by the reflection coefficient from inside the cavity. Here we show how this formula can be derived by using the explicit expressions for the electric ($\bm E$) and magnetic ($\bm B$) fields induced by the emitting electric dipole
\[
{\bm d}(t) = {\bm d}\, {\rm e}^{-{\rm i} \omega t} + {\bm d}^*\, {\rm e}^{{\rm i} \omega t}
\]
inside and outside the spherical cavity. These fields have the following structure 
\begin{eqnarray} \label{el_magn_field_spher}
{\bm E}({\bm r})&=& \left\{ \begin{array}{c} {\bm E}_{{\rm in},1}({\bm r})+
{\bm E}_{{\rm in},2}({\bm r}) \:, \hspace{2 mm} \mbox{if} \hspace{4 mm} r_{\rm QD} < r \leq a\:,\\ 
\hspace{8 mm} {\bm E}_{\rm out}({\bm r}) \:, \hspace{12 mm} \mbox{if} \hspace{3 mm} a \leq r\:, \hspace{1.2 cm} \end{array} \right. \\
{\bm B}({\bm r}) &=& \left\{ \begin{array}{c} {\bm B}_{{\rm in},1}({\bm r})+
{\bm B}_{{\rm in},2}({\bm r}) \:, \hspace{2 mm} \mbox{if} \hspace{4 mm} r_{\rm QD} < r \leq a\:,\\ 
\hspace{8 mm} {\bm B}_{\rm out}({\bm r}) \:, \hspace{12 mm} \mbox{if} \hspace{3 mm} a \leq r\:, \hspace{1.2 cm} \end{array} \right.  
\end{eqnarray}
where 
\begin{eqnarray} \label{el_field_cav}
{\bm E}_{{\rm in},1}({\bm r}) &=& \frac{\rm i}{3} k_0^2 k_1 [ 2 h^{(1)}_0(x_1) {\bm d} + h^{(1)}_2(x_1) \tilde{\bm d}] \:, \\
{\bm E}_{{\rm in},2}({\bm r}) &=& \frac{{\rm i} R^{\rm TM}_{12,1}}{3} k_0^2 k_1 [ 2 j_0(x_1) {\bm d} + j_2(x_1) \tilde{\bm d}] \:, \nonumber\\
{\bm E}_{\rm out}({\bm r}) &=& \frac{{\rm i} T}{3} k_0^2 k_2  [ 2 h^{(1)}_0(x_2) {\bm d} + h^{(1)}_2(x_2) \tilde{\bm d}]
\nonumber
\end{eqnarray}
and
\begin{eqnarray}\label{mag_field_cav}
{\bm B}_{{\rm in},1}({\bm r}) &=& \sqrt{\varepsilon_1} k_0^2 k_1 h^{(1)}_1(x_1) \, {\bm d} \times {\bm n}\:, \\
{\bm B}_{{\rm in},2}({\bm r}) &=& R^{\rm TM}_{12,1} \sqrt{\varepsilon_1} k_0^2 k_1 j_1(x_1) 
\, {\bm d} \times {\bm n}\:, \nonumber\\
{\bm B}_{\rm out}({\bm r}) &=& T\sqrt{\varepsilon_2} k_0^2 k_2 h^{(1)}_1(x_2)\, {\bm d} \times {\bm n}\:.\nonumber
\end{eqnarray}
Here $x_1 = k_1a$, $x_2 = k_2a$, ${\bm n} = {\bm r}/r$, $\tilde{\bm d} = 3 ({\bm d} \cdot {\bm n}){\bm n} - {\bm d}$, the coefficient $R^{\rm TM}_{12,1}$ is introduced in \cite{ajiki}, and $T$ is expressed via the coefficient $T_{12,1}^{\rm TM}$ from the same reference as $T = (\varepsilon_1/\varepsilon_2)T_{12,1}^{\rm TM}$. The coefficients $R^{\rm TM}_{12,1}$ and $T$ in Eqs.~(\ref{el_field_cav}) and (\ref{mag_field_cav}) can be obtained from the boundary conditions at the sphere's surface $r = a$:
\begin{eqnarray}\label{boundary}
x_1 h_0^{(1)}(x_1) - h_1^{(1)}(x_1) - {\rm i} R_{12,1}^{\rm TM} [x_1 j_0(x_1) - j_1(x_1)]  \nonumber \\
=  T [x_2 h_0^{(1)}(x_2) - h_1^{(1)}(x_2)]\:, \nonumber \\
\varepsilon_1 [h_1^{(1)}(x_1) - {\rm i}  R_{12,1}^{\rm TM} j_1(x_1)] = \varepsilon_2 T h_1^{(1)}(x_2)\:.
\end{eqnarray}

It is worth to mention that the Purcell factor can be equivalently presented as the ratio
\begin{equation}
f = \frac{I_{\rm cav}}{I_{\rm bulk}}
\end{equation}
of fluxes of electromagnetic energy radiated by the dipole ${\bm d}$ in a system with the cavity and in the homogeneous material and passing through the sphere of radius $r < a$: 
\[
I = \frac{c r^2}{2 \pi} \int\limits_{4 \pi} d \Omega~{\bm n} \cdot {\rm Re} 
[{\bm E}^*(\bm r) \times  {\bm B}(\bm r)]\:,
\]
where $d \Omega$ is the solid-angle element. Substituting the expressions (\ref{el_field_cav}), (\ref{mag_field_cav}) for the electromagnetic fields and performing the necessary transformations, we obtain Eq.~(\ref{purcell_factor}). Equalizing the energy fluxes at the internal and external boundaries of the sphere of radius $a$ and using the boundary conditions (\ref{boundary}), we arrive at the relation
\[
k_2^{\prime} \left( 1
+ \frac{2 k_2^{\prime \prime}\ |1 - {\rm i} k_2 a|^2}{|k_2|^4 a^3} \right)  {\rm e}^{- 2 k_2^{\prime \prime} a} |T|^2
= k_1 (1 + {\rm Re}\ R_{12,1}^{\rm TM})
\]
between $|T|$ and ${\rm Re}\ R_{12,1}^{\rm TM}$. Here $k_2^{\prime}$ and $k_2^{\prime \prime}$ are the real and imaginary parts of the wave number $k_2$.

In the particular case  $a \ll \delta \ll \lambda$ we approximately have, instead of Eq.~(\ref{boundary}):
\begin{eqnarray}\label{boundary2}
&&x_1 h_1^{(1)}(x_1) -  \frac{2 x_1}{3} R_{12,1}^{\rm TM} =  - T h_1^{(1)}(x_2)\:,\\
&&\mbox{}\varepsilon_1 [h_1^{(1)}(x_1) -   \frac{ x_1}{3} R_{12,1}^{\rm TM}] = \varepsilon_2 T h_1^{(1)}(x_2)\:,\nonumber
\end{eqnarray}
where, in its turn, we can set $h_1^{(1)}(x_j) \approx - {\rm i}/x^2_j$ ($j=1,2$). These equations allow one to estimate the field near the sphere and also find an approximate value of the reflection coefficient
\[
R_{12,1}^{\rm TM} \approx -\frac32 \frac{\mathrm i}{k_1^2a^2} \left( 1 - \frac32 \frac{k_1^2}{k_2^2}\right)\:.
\]
For $k_2 = (1 + {\rm i})/\delta$ we obtain ${\rm Re}\, R_{12,1}^{\rm TM} = (9/8)(\delta^2/k_1a^3)$, in agreement with Eq.~(\ref{thinthick}).

It is of special interest to establish the relation between the Purcell factor defined as the ratio $\tau_{{\rm r},{\rm bulk}}/\tau_{{\rm r},{\rm cav}}$ and the radiation quality factor $Q_{\rm ant}$ defined in the physics of antennas \cite{mclean,ziolkowski} as the ratio of the stored (nonpropagating) energy $W_{\rm nonprop}$ to the flux of radiated energy $I$, or more exactly, as $Q_{\rm ant} = 2 \omega W_{\rm nonprop}/I$. Taking into account that the dimension $r_{\rm QD}$ of the emitter (being the quantum dot radius satisfying the condition $k_1 r_{\rm QD} \ll 1$) is the smallest linear dimension in the cavity system under consideration, we find
\begin{equation} \label{antenna}
Q_{\rm ant} = \frac{1}{f k^3_1 r_{\rm QD}^3 }\:,
\end{equation}
where $f = 1 + {\rm Re}\, R_{12,1}$ is the Purcell factor. For emission into a homogeneous medium, $R_{12,1} = 0$ and $Q_{\rm ant} = (k_1 r_{\rm QD})^{-3}$ in agreement with \cite{mclean}. Ziolkowski and Kipple \cite{ziolkowski} have calculated the radiation quality factor of the antenna modelled by an elementary dipole embedded into the center of the spherical shell of double negative material (i.e., with both negative permittivity and negative permeability). According to Eq.~(\ref{antenna}) the Purcell factor for such (and similar) systems can be found from the equation $f = [(k_1 r_{\rm QD})^3Q_{\rm ant}]^{-1}$ by using the numerically calculated value of $Q_{\rm ant}$. 

In this work a semiconductor quantum dot is considered to play the role of an emitter. Its dielectric response is derived quantum-mechanically, see Eq.~\eqref{polarization}. However, it should be pointed out that the Purcell effect can be obtained solely in terms of the classical mechanics and electrodynamics, for example, for the radiation of an electron executing mechanical oscillating motion in a spherically-symmetric parabolic potential. If the potential minimum is located in the center of a cavity of radius $a$, then according to the second Newton's law of motion we can write
\begin{equation} \label{newton}
m(\ddot{\bm d} + \omega^2_0 {\bm d}) = e [ {\bm E}_{\rm rad. fric.} + {\bm E}_{{\rm in},2}(0)]\:,
\end{equation} 
where $m$ is the electron mass, $\omega_0$ is the resonance frequency of its oscillation, dots mean time derivatives,  $e {\bm E}_{\rm rad. fric.}$ is the force of radiative friction induced by the oscillating charge~\cite{ll2}, and ${\bm E}_{{\rm in},2}(0)$ is the field arising due to existence of the spherical surface and introduced in Eq.~(\ref{el_field_cav}). For the field ${\bm E}_{\rm rad. fric.}$ we have
\begin{equation}  \label{vac}
{\bm E}_{\rm rad. fric.} = \frac{2}{3 c^3} \dddot{\bm d}\:.
\end{equation}
Due to the presence of cavity (for simplicity we take $\varepsilon_1=1$) the total field in the point of oscillating-dipole location has the additional contribution ${\bm E}_{{\rm in},2}(0)$ which can be presented in the form, see Eq.~\eqref{el_field_cav}:
 \begin{equation}
 \label{Refl}
 {\bm E}_{{\rm in},2}(0) = \frac{2 \mathrm i R^{\rm TM}_{12,1}}{3} k_0^3 \bm d\:.
 \end{equation} 
For a weak radiation decay one has $\dddot{\bm d} = - \omega^2_0 \dot{\bm d}$ and Eq.~(\ref{newton}) reduces to
 \begin{equation}
 \label{selftot}
 \ddot{\bm d} + \omega_0^2 \bm d + \frac{2e^2\omega_0^2}{3c^3 m}(1 + R^{\rm TM}_{12,1}) \dot{\bm d} =0\:,
 \end{equation}

from which Eq.~\eqref{purcell_factor} follows.
\section{Summary}
We have developed a theory of the Purcell effect, or the effect of surrounding environment on the emission rate of an electric or magnetic dipole, for spherical cavities of arbitrary size. Special attention has been paid to nonresonant cavities with the radius small in comparison with the wavelength of the dipole radiation. In such systems the enhancement of dipole spontaneous emission rates occurs because of the reflection of emitted electromagnetic wave from the spherical surface of the cavity and 
strong amplification of the field amplitude at the dipole location point. We have obtained asymptotic expressions for the Purcell factor in metallic cavities and showed that the acceleration of emission drastically depends on the relation between the skin depth and the cavity size. 

Experimentally the above effects can be observed in metallic cavities with small holes which allow the radiation to escape outside the cavity. The emission acceleration can be also studied in experiments on the structures where the emitter (a molecule or a localized exciton) is located near but outside a metallic particle~\cite{koenderink,anger,kuhn,toropov,gk1,gk2,gk3}.

\acknowledgements
The financial support from RAS, RFBR, Russian President grant for young scientists, and Dynasty Foundation--ICFPM is acknowledged. GK would like to acknowledge support from NSF AMOP (PHY-0757707) and AFOSR (FA9550-10-1-0003).


\begin{thebibliography}{16}
\bibitem{Purcell} E.M. Purcell, Phys. Rev. {\bf 69}, 681 (1946).

\bibitem{oraevskii} A.N. Oraevskii, Usp. Fiz. Nauk {\bf 164}, 415 (1994) [Physics-Uspekhi {\bf 37}, 393 (1994)].

\bibitem{Ivchenkobook2} E.L.~Ivchenko, \textit{ Optical
Spectroscopy of Semiconductor Nanostructures} (Alpha Science Int.,
Harrow, UK, 2005).
\bibitem{Ivch_1992} E.L. Ivchenko and A.V. Kavokin, Fizika Tverdogo Tela {\bf 34}, 1815 (1992) [Sov. Phys. Solid State {\bf 34}, 968 (1992)].

\bibitem{Kalit} M.A. Kaliteevski, S. Brand, R.A. Abram, V.V. Nikolaev, M.V. Maximov, C.M. Sotomayor Torres,
A.V. Kavokin, Phys. Rev. B {\bf 64}, 115305 (2001). 
\bibitem{ajiki} H. Ajiki, T. Tsuji, K. Kawano, K. Cho, Phys. Rev. B {\bf 66}, 245322 (2002).
\bibitem{coccioli} R. Coccioli, M. Boroditsky, K.W. Kim, Y. Rahmat-Saii, E. Yablonovitch, 
IEE Proc.-Optoelectronics {\bf 145}, 391 (1998).
\bibitem{andreani} L.C. Andreani, G. Panzarini, J.M. G\'erard, Phys. Rev. B {\bf 60}, 13 276 (1999).
\bibitem{robinson}J.T. Robinson, C. Manolatou, Long Chen, M. Lipson, Phys. Rev. Lett. {\bf 95}, 143901 (2005).
\bibitem{koenderink} A.F. Koenderink, Optics Lett. {\bf 35}, 4208 (2010).
\bibitem{IEEE} {S. Gallagher and W.J. Gallagher, IEEE Trans. Nuclear Science {\bf NS-32}, 2980 (1985).}
\bibitem{ll8} { L.D. Landau and E.M. Lifshitz, \textit{Electrodynamics of Continuous Media, Second Edition} (Butterworth-Heinemann, Oxford, 2004).} 
\bibitem{ohtaka} K. Ohtaka, A.A. Lukas, Phys. Rev. B {\bf 18}, 4643 (1978).
\bibitem{Yablon} E. Yablonovitch, T.J. Gmitter, R. Bhat, Phys. Rev. Lett. {\bf 61},
2546 (1988).
\bibitem{glauber} R.J. Glauber, M. Lewenstein, Phys. Rev. A {\bf 43}, 467 (1991).
\bibitem{Lavallard} P. Lavallard, M. Rosenbauer, T. Gacoin, Phys. Rev. A {\bf 54}, 5450 (1996).
\bibitem{khitrova1} A. Thr\"anhardt, C. Ell, G. Khitrova, H. M. Gibbs, Phys. Rev. B {\bf 65}, 035327 (2002).
\bibitem{GoupPRB} S.V. Goupalov, Phys. Rev. B {\bf 68}, 125311 (2003).
\bibitem{mclean} J.S. McLean, IEEE Trans. Antennas Propagat. {\bf 44}, 672 (1996).
\bibitem{ziolkowski} R.W. Ziolkowski, A.D. Kipple, IEEE Trans. Antennas Propagat. {\bf 51}, 2626 (2003).
\bibitem{ll2} L.D. Landau and E.M. Lifshitz, \textit{The classical theory of fields} (Pergamon Press, London, 1987).
\bibitem{anger} P. Anger, P. Bharadwaj, L. Novotny, Phys. Rev. Lett. {\bf 96}, 113002 (2006). 
\bibitem{kuhn} S. K\"{u}hn, U, H\aa kanson, L. Rogobete, V. Sandoghdar, Phys. Rev. Lett.  {\bf 97}, 017402 (2006).
\bibitem{toropov} A. A. Toropov, T. V. Shubina, V. N. Jmerik, S. V. Ivanov, Y. Ogawa, F. Minami, Phys. Rev. Lett. {\bf 103}, 037403 (2009).
\bibitem{gk1} A. G. Curto, G. Volpe, T. H. Taminiau, M. P. Kreuzer, R. Quidant,  N. F. Van Hulst, Science {\bf 329}, 930 (2010).

\bibitem{gk2} N. Meinzer, M. Ruther, S. Linden, C. M. Soukoulis, G. Khitrova, J. Hendrickson, J. D. Olitzky, H. M. Gibbs,  M. Wegener,  Opt. Express {\bf 18},  24140 (2010).

\bibitem{gk3} K. Tanaka, E. Plum, J. Y. Ou, T. Uchino,  N. I. Zheludev, Phys. Rev. Lett. {\bf 105}, 227403 (2010). 
\end{thebibliography}
\end{document}